\documentclass[10pt,a4paper]{article}
\usepackage{geometry}
\usepackage{bbm}
\usepackage{amsmath}
\usepackage{natbib}

\usepackage{dcolumn}
\usepackage{graphicx}

\usepackage{bbm}
\usepackage{amsmath}
\usepackage{graphicx}
\usepackage{dcolumn}
\usepackage{multirow}
\usepackage{subcaption}
\usepackage{changepage}
\usepackage[]{placeins}
\usepackage{setspace}
\usepackage[]{appendix}

\usepackage[textsize=scriptsize, disable]{todonotes}

\usepackage{array}
\newcolumntype{L}[1]{>{\raggedright\let\newline\\\arraybackslash\hspace{0pt}}m{#1}}
\newcolumntype{C}[1]{>{\centering\let\newline\\\arraybackslash\hspace{0pt}}m{#1}}
\newcolumntype{R}[1]{>{\raggedleft\let\newline\\\arraybackslash\hspace{0pt}}m{#1}}

\usepackage[textsize=scriptsize]{todonotes}

\graphicspath{{"C:/Users/epo21/Dropbox/Electricity_Demand_Causal_Tree_Project/Papers/WorkingPaper/graphs/"}}

\newcommand{\pkg}[1]{\texttt{#1}}

\begin{document}
	
\appendixtitleon
\appendixtitletocon	
\thispagestyle{empty}
\title{Causal Forest Estimation of Heterogeneous Household Response to Time-Of-Use Electricity Pricing Schemes}
\author{Eoghan O'Neill \\
Faculty of Economics  \\
University of Cambridge \smallskip \\
\and Melvyn Weeks\thanks{Contact Author: Dr. M. Weeks, Faculty of Economics,
University of Cambridge, Cambridge
CB3 9DD, UK. Email: mw217@econ.cam.ac.uk. Our thanks are due to Kai Liu, David Newbery, Alexei Onatski, and Michael Pollitt.
} \\
Faculty of Economics and Clare College, \smallskip \\ University of Cambridge}
\maketitle
	
	\begin{abstract}
\small

We examine the household-specific effects of the introduction of Time-of-Use (\textsc{tou}) electricity pricing schemes. Using a causal forest \citep{athey2016recursive, wager2018estimation,athey2019generalized}, we consider the association between past consumption and survey variables, and the effect of \textsc{tou} pricing on household electricity demand. We describe the heterogeneity in household variables across quartiles of estimated demand response and utilise variable importance measures. 

Household-specific estimates produced by a causal forest exhibit reasonable associations with covariates. For example, households that are younger, more educated, and that consume more electricity, are predicted to respond more to a new pricing scheme.  In addition, variable importance measures suggest that some aspects of past consumption information may be more useful than survey information in producing these estimates. 

\medskip



		
\end{abstract}

	\newpage
	
	
	\newpage



\pagebreak
	
	\section{Introduction}

If a policymaker believes the impacts of a particular policy are heterogeneous in a given population, then it is helpful to describe the distribution of household-specific effects of the policy. The critical question is: does the policymaker know ex ante which characteristics of individuals are driving the differences in the impact of the policy? 

Researchers often describe subpopulations that are of interest a priori, and which can be defined by a known combination of covariates. However, increasingly researchers have many covariates at their disposal and it may not be clear which covariates should be used to categorise heterogeneity, nor what functional form best describes the association between these covariates and treatment effects.  

The introduction of an electricity pricing scheme is an example of a policy with heterogeneous effects. Consumers in different socioeconomic groups and with distinct historical intra-day load profiles and behavioural characteristics, may respond differently to the introduction of
tariffs that charge different prices for electricity at different times of the day. Customers who can (cannot) adapt their consumption profile to \textsc{tou} tariffs will accrue a benefit (cost). Those who consume
electricity at more expensive peak periods, and who are unable to change their consumption
patterns, could end up paying significantly more.

In assessing whether demographic variables are informative in terms of
the impact of \textsc{tou} tariffs on load profiles, the Customer-Led Network Revolution project \citep{sidebotham2015customer} noted
\begin{quote}
.. a relatively consistent average demand profile across the different demographic
groups, with much higher variability \emph{within} groups than \emph{between} them. This high
variability is seen both in total consumption and in peak demand.
\end{quote}
In addition, the question of which demographic variables are important when considering the impact
of energy policies ignores the fact that many of these variables should be considered together, in
a multiplicative fashion. One reason for this finding might be that it is the (unknown) combination of income, household size, education, and daily usage patterns that describes a particularly responsive or unresponsive group.

	In this paper we consider the household-specific effects on customers following the introduction of a Time-of-Use (\textsc{tou}) pricing scheme where the price per kWh of electricity usage depends on the time of consumption. The pricing scheme is enabled by smart meters, which records consumption every half-hour. Using machine learning methods, we describe the association between the effect of \textsc{tou} pricing schemes on household electricity demand and a range of variables that are observable before the introduction of the new pricing schemes. 

We apply a recently developed method, known as a causal forest, which aggregates over estimates from causal trees \citep{athey2016recursive, wager2018estimation,athey2019generalized}. This method searches across covariates for good predictors of heterogeneous treatment effects.  A causal tree provides an interpretable description of heterogeneity, and causal forests can be used to obtain individual-specific estimates of treatment effects. Heterogeneous effects are described by Conditional Average Treatment Effect (\textsc{cate}) estimates, which are the expected effects of a treatment for individuals in subpopulations defined by covariates. We characterize the most and least responsive households by applying the methods described by \cite{chernozhukov2017generic}.

Given that policy makers are often interested in the factors underlying a given prediction, it is desirable to gain some insight to which variables in the large set of covariates are most often selected. A key challenge follows from that fact that partitions generated by tree-based methods are sensitive to subsampling, whereas the use of an ensemble method such as causal forests produces more stable, but less interpretable estimates.

To address this problem we utilise variable importance measures to consider which variables are chosen most often by the causal forest algorithm. However, in the estimation of variable importance it is important to account for the impact of the varying information content across continuous versus discrete random variables. In particular, tree based methods can be biased towards continuous variables, given the presence of more potential splitting points. We address this issue by including permutation-based tests for our variable importance results. This is particularly important for this analysis given that many of our demographic variables are either binary or categorical.  


In section \ref{section_methods} we first describe the potential outcomes framework and conditional average treatment effects, then describe causal trees and causal forests.  We describe the variable importance measures and outline how we will apply the methods of \cite{chernozhukov2017generic} to describe heterogeneity between the most and least demand responsive households. In section \ref{het_hhold_response}, we introduce the application to electricity smart meter data, and review existing literature. In section \ref{results}, we present the results. Section \ref{conclusion} concludes.


\section{Methods for Estimation of Heterogeneous Treatment Effects}\label{section_methods}
		

The estimand is defined using the potential outcomes framework introduced by \cite{neyman1923applications} and developed by \cite{rubin1974estimating}. Let $X_i$ be a vector of covariates for individual $i$. Suppose that there is one treatment group of interest. $Y_i (1)$ ($Y_i (0)$) denotes the potential outcome if individual $i$ is allocated to the treatment (control) group.  The causal effect of a treatment on individual $i$ is therefore $Y_i (1) - Y_i (0)$. The fundamental problem of causal inference is that we do not observe the causal effect for any $i$ \citep{holland1986statistics}.

The estimand that we consider is the Conditional Average Treatment Effect (\textsc{cate})
\begin{equation}\label{CATE_eq}
\tau (x)= E[Y_i(1)-Y_i(0) | X_i = x].
\end{equation}
Whereas the \textsc{ate} can be estimated by a difference in means $\bar{y}_t - \bar{y}_c$, where $\bar{y}_t$ ($\bar{y}_c$) is the mean of the outcome variable for the treated (control) group, the \textsc{cate} can be thought of as a subpopulation average treatment effect.\footnote{In instances where we condition on $x$ being in some subset of the covariate space, i.e. $x \in \mathbbm{A} \subset \mathbbm{X} $, and  $\tau_{\mathbbm{A}}= E[Y_i(1)-Y_i(0) | x \in  \mathbbm{A}] $, we also refer to this as the \textsc{cate} (with suitably re-defined covariates). } \footnote{Another estimand is the average treatment effect conditional upon observed covariates $\bar{\tau} = \frac{1}{N} \sum_{i=1}^{N} \tau(x_i) = \frac{1}{N} \sum_{i=1}^{N} E[Y_i(1)-Y_i(0) | X_i = x_i]$. \cite{imbens2015causal} refer to this as the conditional average treatment effect, but we shall use the above definition of the \textsc{cate}.
} The \textsc{cate} is identified under unconfoundedness, i.e. $Y_i(1), Y_i(0) \perp T_i | X_i $ , and overlap, i.e. $0 < \Pr (T_i = 1 | X_i = x) <1 \ \forall \ x$, where $T_i$ denotes the treatment indicator variable.

A \textsc{cate} estimate can be obtained from a linear model by including interactions between the treatment indicators and the conditioning variable(s) of interest.
The inclusion of interaction terms in a linear model is a common technique for exploring the heterogeneity of treatment effects in areas ranging from biomedical science to the social sciences.\footnote{A description of the application of linear regression methods for the purpose of estimating treatment effects in randomized experiments can be found in \cite{athey2017econometrics}.}


It is possible to search for heterogeneity in treatment effects simply by separately estimating \textsc{cate}s using many possible conditioning variables and repeatedly estimating the standard linear regression model, and conducting tests of multiple hypotheses. However, a clear problem is false discovery and the need to adjust significance levels for multiple hypothesis testing which can limit the power of a test to find heterogeneity.

A number of alternative machine learning methods allow the researcher to explore more complex forms of heterogeneity. Recent methods for \textsc{ite} estimation include \textsc{lasso} \citep{imai2013estimating, weisberg2015post,tian2014simple}, \textsc{bart} \citep{hill2011bayesian, hahn2017bayesian, logan2019decision}, other tree-based methods \citep{powers2017some,oprescu2018orthogonal, lu2018estimating,lechner2019modified}, the R-learner \citep{nie2017quasi}, neural networks \citep{shalit2017estimating, farrell2018deep, atan2018deep, shi2019adapting}, Generalized Adversarial Networks \citep{yoon2018ganite}, and many more. 

In this study we are interested in allowing for many possibly nonlinear interactions between covariates. Forest methods perform well in capturing nonlinear interactions. Furthermore causal forests perform reasonably well relative to other methods and have known asymptotic properties \citep{knaus2018machine, alaa2019validating, athey2019generalized}. Therefore we apply the causal forest method for \textsc{ite} estimation.

\subsection*{Regression and Causal Trees}

Causal forests \citep{wager2018estimation,athey2019generalized} average the predictions of many causal trees \citep{athey2016recursive}. Causal trees are decision trees for treatment effect estimation, and can be viewed as a variation on standard regression trees, with a different splitting criterion, and different terminal node estimates.

A single regression tree is constructed as follows \citep{friedman2009elements,breiman1984classification}. Suppose there are $p$ covariates and $N$ observations. The objective is to partition the covariate space $\mathbbm{X}$  into $M$ mutually exclusive regions $R_1,...,R_M$, where the outcome for an individual with covariate vector $x$ in region $R_m$ is estimated as the mean of the outcomes for training observations in leaf $R_m$. The following algorithm is used to apply binary splits of the data:

\medskip

Let $X_j$ be a splitting variable and $s$ be a split point. Define $R_1(j,s) = \{X | X_j \le s \} $ and $R_2 (j,s) = \{ X | X_j > s \}$.\footnote{	If a splitting variable is categorical with $q$ unordered values, then we can consider all $2^{q-1}-1$ possible splits of the $q$ values into two groups, or we can use binary variables for each category.} The algorithm selects the pair $(j,s)$ that solves:
\begin{equation}\label{CART_criterion}
\min_{j,s} \left[  \sum_{x_i \in R_1(j,s) } (y_i - \bar{y}_1(j,s) )^2 +  \sum_{x_i \in R_2 (j,s)} (y_i - \bar{y}_2(j,s) )^2  \right]
\end{equation}
where  $\bar{y}_1(j,s)$ and $\bar{y}_2(j,s)$  are the mean outcomes in $R_1(j,s)$ and $R_2(j,s)$ respectively.
When the data has been split into two regions, the same process is applied separately to each region. Then the process is repeated on each of the four resulting regions, and so on.

Trees can be fully grown, or grown up to a stopping rule, or a penalty term can be included in the splitting criterion that penalizes the tree size \citep{friedman2009elements}. Causal tree \citep{athey2016recursive} leaf estimates are differences in means between treated and untreated observations, and the splitting criterion is different to (\ref{CART_criterion}) because the goal is to minimize the expected mean square error of these treatment effect estimates. 


\cite{athey2016recursive} also note that estimates produced by standard regression tree algorithms are biased because the same data is used for tree construction and for estimating the terminal node means. \cite{athey2016recursive} therefore suggest separating the training data into two independent subsamples, one for construction of the tree, and one for estimation of the terminal node means. This so-called \textit{honest} estimation ensures unbiased estimates.

\subsection*{Random and Causal Forests}
		
The prediction of a random forest \citep{friedman2009elements} is the average of many ($B$) unpruned regression trees. Each tree, $T_b$ ($b$ indexing the bootstrap samples), is produced using a bootstrap sample of size $N$  without replacement from the training data. At each split in the tree, the algorithm uses a random subset of the set of all covariates as potential splitting variables. Each tree is fully grown up to a minimum leaf size. 


The prediction for an individual with a vector of covariates $x$ is then $\frac{1}{B} \sum_{b=1}^{B} T_b (x)$, where $T_b (x)$ is the estimate produced by tree $b$.
The trees are not independent because two bootstrap samples can have some common observations, and therefore the correlation between trees limits the benefits of averaging. However, this correlation is reduced through the random selection of the input variables.

Similar aggregations over causal trees, known as causal forests, can improve the accuracy of treatment effect estimates. \cite{wager2018estimation} outline the properties of causal forests and show that, under certain assumptions, the predictions from causal forests are asymptotically normal and centred on the true treatment effect for each individual. Recent applications of causal forests can be found in articles by \cite{davis2017rethinking, davis2017using} and \cite{bertrandcontemporaneous}. 

\cite{athey2019generalized} introduce a generalization of random forests which can be viewed as an adaptive kernel method. This generalized random forest (\textsc{grf}) framework can be used for estimation of a variety of models, including treatment effect estimation. The causal forest method introduced by \cite{wager2018estimation} is almost equivalent \todo{\tiny This is vague (\emph{almost}. Please add footnote to explain. Is this equivalence only true for linear models based on CATE? W \& Athey (2019) makes this point} \todo[inline, color=yellow]{``Almost equivalent'' is the phrase used on page 22 of the GRF paper (\emph{so it should be referenced or minimally italicised.} The equivalence is exact for standard RF (hence ``GRF''). I don't understand the linear model question, what is the original ML method for which grf has an approximation? (grf provides an exact implementation of RF, and approximate implementation of standard CF).}to the \textsc{grf} implementation of a causal forest without centering. \textsc{grf} involves an approximate, gradient-based loss criterion, and orthogonalizes the outcome treatment variables from estimates produced by separate forests before fitting the causal forest.

\subsection*{Variable Importance}			

A general issue which applies to standard regression trees and random forests is the trade-off between interpretability and stability.  A single causal tree splits the data into relatively few leaves. The results are easy to interpret given that a simple tree diagram allows the researcher to quickly identify the subgroup to which any household belongs by following a set of decision rules.  \cite{breiman2001random} and \cite{strobl2008statistical} note that single trees can be unstable with small changes in the training data resulting in a very different model (tree). However,  although  stable forests generate better predictive performance, the interpretability of a single tree is lost when we move to an ensemble method, such as a causal forest.

 Across the many trees within a forest, it is not immediately clear what covariates most strongly influence the final estimates, and how different covariates interact. 
 Variable importance measures describe which variables are chosen most often by the causal forest algorithm. However, in the estimation of variable importance it is important to account for the impact of the varying information content across continuous versus discrete random variables. In particular, tree based methods can be biased towards continuous variables, given the presence of more potential splitting points. We address this issue by including permutation-based tests for our variable importance results.
We apply the  default variable importance measure for the command \pkg{causal\_forest} in the \pkg{R} package \pkg{grf}. This variable importance measure is based upon a count of the proportion of splits on the variable of interest up to a depth of 4, with a depth-specific weighting.
\footnote{Variable importances for categorical variables are the sum of the variable importances of binary variables. The parameters we set for the \pkg{causal\_forest} command are: 15000 trees,  bootstrap samples of half the data,  one third of covariates randomly drawn as potential splitting variables, and minimum node size of 5.}
\begin{equation}
imp(x_j) = \frac{ \sum_{k=1}^{4} \left[ \frac{\sum_{all \ trees} number \ depth \ k \ splits \ on \ x_j }{\sum_{all \ trees} total \ number \ depth \ k \ splits}  \right] k^{-2}   }{\sum_{k=1}^{4} k^{-2}}		
\end{equation}

\subsection*{Permutation Test for Causal Forest Variable Importance}

If the splits in trees spuriously occur (in the sense that variables might not be as important, or strongly associated with the outcome, as suggested by the number of splits) more often on continuous variables and variables with more categories, then this should also occur when the dependent variable is permuted. In this instance, the p-value should be unaffected unless the extent of the over-selection of variables for splitting is also dependent on the true importance of the variables. We investigate this issue in further detail in Appendix \ref{simstudy_perm_test}, which contains a simple simulation study of this permutation based variable importance test. The simulations suggest that the p-values are potentially unaffected by the bias of variable splitting towards variables with more possible splitting points.


Following the method of \cite{altmann2010permutation} for random forests,\footnote{\cite{altmann2010permutation} show that p-values based on permutation of the dependent variable can address the issues of bias towards variables with more categories, and masking of the importance of groups of highly correlated variables.} and \cite{bleich2014variable} for BART, we compute p-values for the default variable importances provided by the \pkg{grf} package. This involves permuting the dependent variable 1000 times and obtaining variable importances for all variables from 1000 causal forests fitted separately using the 1000 permutations as dependent variables. The variable importances are also obtained from a causal forest using the original, unpermuted dependent variable. Then, following the ``local'' test described by \cite{bleich2014variable}, we obtain a p-value for each variable by  finding the proportion of the 1000 causal forests for which the variable had a greater variable importance measure than that obtained from the causal forest with the unpermuted dependent variable.

\subsection*{Testing and Describing Treatment Effect Heterogeneity}

In the results section we provide point \textsc{ite} estimates with confidence intervals obtained using the methods described by \cite{athey2019generalized}. However, we are also interested in describing heterogeneity through \textit{features} of the treatment effect function $\tau(x)$ (Equation \ref{CATE_eq}), which requires a different approach to inference \citep{chernozhukov2017generic} involving repeatedly obtaining two random subsamples, and training a causal forest on one subsample, and performing a statistical test of interest on the other subsample. A sample split allows for valid inference conditional on the subsample of data used for constructing the causal forest, and repeated sample splitting is used in accounting for the uncertainty induced by the random sampling. This requires the training of many causal forests, in contrast to the requirement of a single causal forest for valid \textsc{ite} prediction intervals.

We apply the methods of \cite{chernozhukov2017generic} to first test for the presence of heterogeneity, and then characterize the association between covariates and demand response. This approach, summarized below, involves repeated data splitting to avoid overfitting and to achieve validity.

Let $Y$ be the outcome variable, $D$ be the treatment indicator variable, and $Z$ be all other covariates. We split the data in half into a main sample $Data_M$ and auxiliary sample $Data_A$ 1000 times. For each split we train a causal forest on $Data_A$ and also a regression forest on the untreated observations in $Data_A$. Then we obtain treatment effect estimates, $S(Z)$ by applying the trained causal forest to $Data_M$, and we obtain baseline outcome estimates, $B(Z)$ by applying the trained regression forest to $Data_M$.\footnote{$B(Z)$ is included to improve efficiency. Inference would still be valid if we removed $B(Z)$ from equations \ref{BLP_eq} and \ref{GATEs_eq}.} This will result in 1000 sets of parameter estimates that can be used for valid inference on the parameters. See below for a description of the parameters of interest, and see \cite{chernozhukov2017generic} for a description of the inference methods. This approach accounts for estimation uncertainty conditional on the auxiliary sample and splitting uncertainty induced by random partitioning of the data into main and auxiliary samples. \citep{chernozhukov2017generic}.

First, we test for heterogeneity using the Best Linear Predictor (\textsc{blp}) of the \textsc{cate} \citep{chernozhukov2017generic}. We obtain the following estimated model by weighted \textsc{OLS}:
\begin{equation}
Y_i = \hat{\alpha_0} + \hat{\alpha_1}B(Z_i)+\hat{\beta_1}(D_i-p(Z_i))+\hat{\beta_2}(D_i-p(Z_i))(S(Z_i)-\overline{S(Z)})+\hat{\varepsilon}_i
\label{BLP_eq}
\end{equation}
where the weights are $\{\hat{p}(Z)(1-\hat{p}(Z))\}^{-1}$ and $\overline{S(Z)}=|M|^{-1}\sum_{i \in M} S(Z_i)$. For the randomized controlled trial dataset used in this paper, we set $\hat{p}(Z)$ equal to the sample proportion of treated individuals.

$\beta_2$ reflects the extent to which the estimated treatment effect is a proxy for the true treatment effect function (\ref{CATE_eq}). Rejection of the null hypothesis $\beta_2=0$ implies that there is heterogeneity and $S(Z)$ is a relevant predictor. \cite{chernozhukov2017generic} outline how to perform valid inference on $\beta_2$. For each of the 1000 data splits into main and auxiliary samples, we keep the estimates $\hat{\beta}_1$, $\hat{\beta}_2$, and upper and lower bounds of 95\% confidence intervals. The medians of the $\hat{\beta}_1$ and $\hat{\beta}_2$ are the final $\beta_1$ and $\beta_2$ estimates. Similarly, medians of upper and lower bounds define the confidence intervals.\footnote{The final upper bound is the lower median of the 1000 upper bounds, and the final lower bound is the upper median of the 1000 lower bounds. The final estimates of $\beta_1$ and $\beta_2$ are mid-points of lower and upper medians  of $\hat{\beta}_1$ and $\hat{\beta}_2$ \citep{chernozhukov2017generic}.} The confidence level of the final interval is $90\%$, and accounts for splitting uncertainty.

We use the estimated treatment effects $S(Z)$ to divide the main sample into groups $G_1$ to $G_4$, where $G_1$ is the 25\% of the data that has the lowest (i.e. most negative) treatment effect estimates and $G_4$ has the highest treatment effect estimates. Sorted Group Average Treatment Effect (\textsc{gate}) estimates \citep{chernozhukov2017generic} are obtained from the following estimated model:
\begin{equation}Y_i = \hat{\alpha}_0 + \hat{\alpha}_1 B(Z_i) + \sum_{k=1}^K \hat{\gamma}_k \mathbbm{I}(G_k)+\hat{\epsilon}_i
\label{GATEs_eq}
\end{equation}
where $\mathbbm{I}(G_k)=1$ if individual $i$ is in group $G_k$ and 0 otherwise. The weights are the same as in equation \ref{BLP_eq}. Inference on $\gamma_k$ and the difference $\gamma_4-\gamma_1$ is made using the same approach as for $\beta_2$ in (\ref{BLP_eq}).  

\cite{chernozhukov2017generic} also outline how to perform valid inference on the average of any function of the outcome and pre-trial covariates, $g(Y,Z)$, within group $G_k$, and differences in these averages between groups $G_1$ and $G_4$. This is referred to as Classification Analysis (\textsc{clan}) and we utilise this method to test for differences in the outcome and pre-trial covariates between the most and least affected 25\%.

In summary, the methods of \cite{chernozhukov2017generic} allow us to test for the existence of heterogeneity; test the relevance of our \textsc{ite} estimates; and  to characterise heterogeneity in the treatment effects by describing the most and least affected individuals.

\section{Heterogeneity of Household Electricity Demand Response}\label{het_hhold_response}	


\textsc{tou} tariffs are becoming more implementable through the roll-out of smart metering technology. The subsequent increase in the availability of large amounts of past electricity consumption data allows for more household specific targeting of electricity pricing and other demand stimuli. Furthermore, in a world where energy suppliers rely increasingly on renewables which are intermittent in nature, measures to reduce peak demand are required as part of the need to balance supply and demand. Understanding  heterogeneity in household responses to \textsc{tou} pricing is of interest to both regulators and retailers.

The British energy regulator, \cite{ofgem2013}, is interested in the impact of new pricing schemes upon vulnerable and low income customers.
\cite{faruqui2010impact} postulate that two potentially offsetting forces influence how we expect low-income customers to be impacted differently by new electricity pricing schemes. First, lower income customers can have a greater proportion of their demand in off-peak hours, and therefore can benefit from \textsc{tou} pricing without adjusting their daily demand profile. Second, we might not expect these customers to shift and reduce load as much as other customers because they have lower usage levels in general and less discretionary usage. \cite{faruqui2010impact} confirm these hypotheses using US data, and find that low income customers change their electricity usage less than higher income customers.


Counter to some of this evidence, studies by Lower Carbon London \citep{schofield2014} and Frontier Economics and Sustainability First \citep{decc2012} have noted the generally low associations between demographic variables and demand response, and in particular, the lack of evidence pertaining to differing responses of low-income and vulnerable customers. One possible reason for this is that individuals most affected by energy policies might be identified through the interaction of a number of variables. For example, the Centre for Sustainable Energy produced a report \citep{preston2013hardest} which used interactions of variables to define the groups of households predicted to face the largest increase in household bills as a result of changes in energy policy.

In this study we examine the importance of variables constructed from historic load profiles. Relatively few studies have conditioned upon past usage data when estimating treatment effects of electricity pricing schemes. Some recent examples include a study using US data by \cite{harding2016empowering}, who split the sample into low, medium, and high usage customers. The results suggest that high usage customers decrease peak usage to a greater extent, which is expected because these customers have more reducible usage. However, surprisingly low-income customers appear to increase consumption in off-peak time periods. The authors note that this substantial load-shifting by low-income customers demonstrates the difficulty in anticipating the impact of new pricing schemes for some customer segments. A number of recent studies have used past electricity usage data for the estimation of household-specific treatment effects. \cite{bollinger2015welfare} condition upon the empirical distribution of past electricity usage and consider how a utility can gain from targeting based upon \textsc{ite} estimates. \cite{balandat2016new} estimates \textsc{ite}s by comparing predictions of electricity usage under control group allocation to realised usage under treatment allocation during the trial period.

\subsection*{Data}\label{data}		

The dataset used in this project is from the Electricity Smart Metering Customer Behavioural Trial conducted by the Irish Commission for Energy Regulation \citep{cer2011}. The \textsc{cer} note that this is ``one of the largest and most statistically robust smart metering behavioural trials conducted internationally to date'' \citep{cer2011}.
The dataset consists of half hourly residential electricity demand observations for 4225 households over 536 days. The benchmark period began on 14th July 2009 and ended on 31st December 2009. Households were then randomly allocated to either a control group or various \textsc{tou} Pricing Schemes and Demand Side Management stimuli from 1st January 2010 to 31st December 2010.

\begin{figure}
	\centering
	\begin{subfigure}{\linewidth}
		\includegraphics[width=\linewidth]{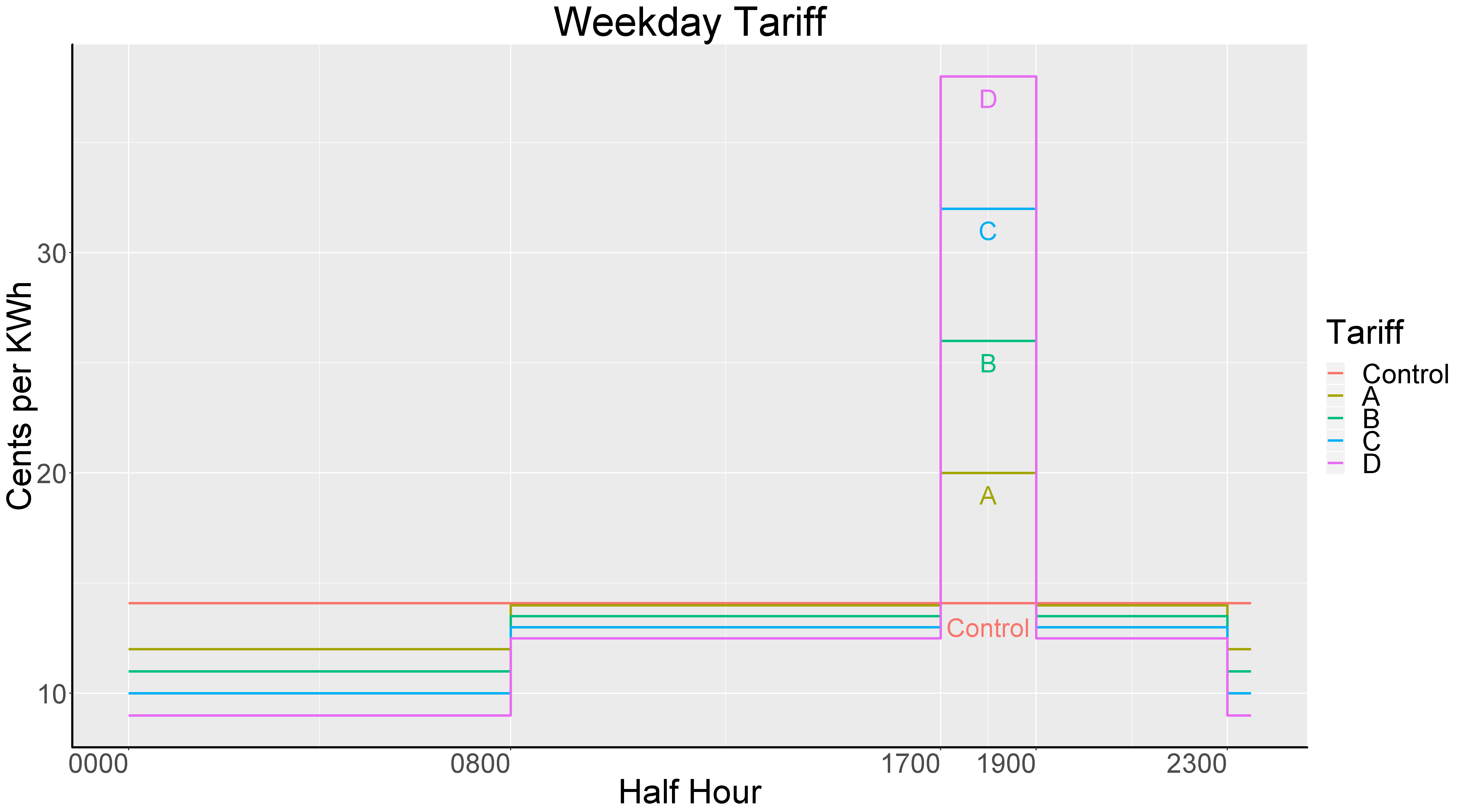}
		\caption{Trial period \textsc{tou} tariffs}
		\label{TrialTariffs}
	\end{subfigure}
	\begin{subfigure}{\linewidth}
		\includegraphics[width=\linewidth]{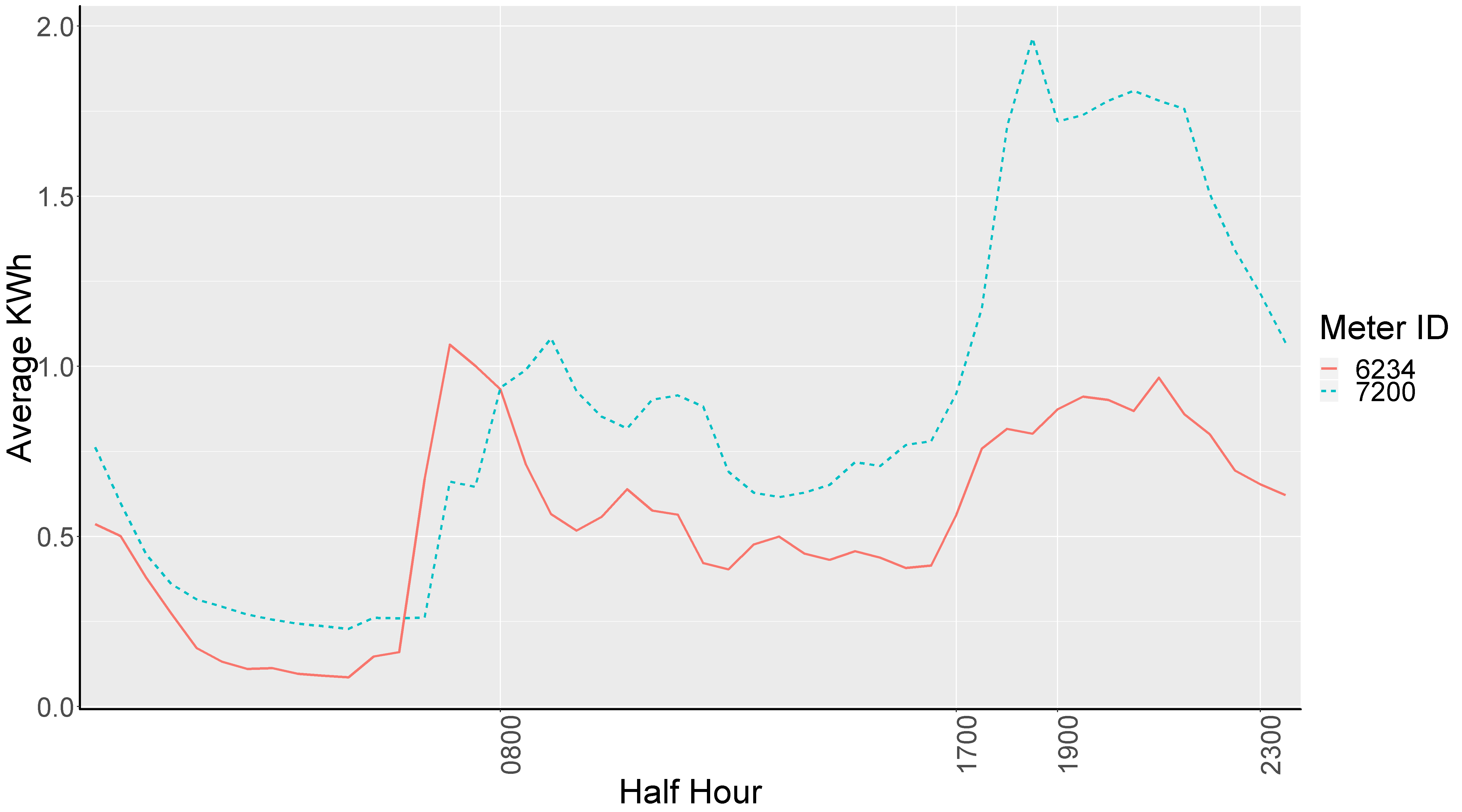}
		\caption{Pre-trial average half-hourly demand for two households}
		\label{hhold_profile_example1}
	\end{subfigure}
	\caption{Prices and examples of demand profiles}
	\label{prices_profiles}
\end{figure}

All households were charged the normal Electric Ireland tariff of 14.1 cents per kWh during the benchmark period. During the trial period the control group remained on the tariff of 14.1 cents per kWh while the test group were allocated to tariffs \textsc{a}, \textsc{b}, \textsc{c}, or \textsc{d}.\footnote{There was also a Weekend tariff group, which we exclude from this study.} The tariffs \textsc{a} to \textsc{d} were structured as shown in Table \ref{tariff_table}, and are graphed in Figure \ref{TrialTariffs}.


Households in the test group were also allocated to one of the following Demand Side Management (\textsc{dsm}) stimuli: Bi-monthly detailed Bill; Monthly detailed bill; Bi-monthly detailed bill and In-Home Display (\textsc{ihd}); Bi-monthly detailed bill and Overall Load Reduction (\textsc{olr}) incentive.

The identification of \textsc{ate}s depends upon unconfoundedness and overlap. The \textsc{cer} took a number of steps to ensure that the samples for treatment groups were representative and did not exhibit notable biases.
A stratified random sampling framework was used with phased recruitment. Non-respondents and attriters were surveyed and adjustments were made accordingly. Those who opted in were compared to the national profile.
The full dataset contains 4225 households, with 768 households in the control group and 233 households facing the combination of tariff \textsc{c} and \textsc{ihd} stimulus, which will be the treatment group of interest in this study.

Figure \ref{hhold_profile_example1} gives an example of average half hourly usage on weekdays before the trial period for households with similar survey responses. The two households both have four people in a 3 bedroom semi-detached house, in which the chief earner is an employee and lower middle class with 3rd level education. Both households also typically have one person at home during the day, own their home, have timed oil heating, and have a similar stock of appliances. This figure shows that even households that are similar across multiple characteristics do not necessarily have the same patterns of demand use.  Therefore these type of survey variables are of limited use in describing demand heterogeneity.\footnote{We make use of pre-trial survey data, but we cautiously avoid using post-trial survey information. \cite{prest2017peaking} applies a causal tree method to this data, but the estimates are potentially biased by conditioning on post-trial survey information. Our methods also differ from those of \cite{prest2017peaking} in that we use a causal forest.} 



	\section{Results}\label{results}
	
	 The outcome variable is average half-hourly peak time electricity consumption during the trial period (measured in kWh), excluding weekends. We restrict attention to Tariff \textsc{c} in combination with the In-Home Display (\textsc{ihd}). The \textsc{ihd} stimulus is of greater interest than the other information stimuli, and tariff \textsc{c} has a high ratio of peak to off-peak prices and more observations than any other tariff combined with the \textsc{ihd}.\footnote{343 households were allocated to Tariff C with the IHD, whereas only 126 households were allocated to tariff D with the IHD.} 

	 The standard \textsc{ate} estimates for the tariff \textsc{c} with \textsc{ihd} range from -0.073 to -0.092 kWh for an average peak half hour, depending on the set of controls.\footnote{These results are obtained by linear regression of average peak usage on the treatment indicator.} 
	Mean half-hourly peak consumption for the control group during the trial period (one full year) was 0.799 kWh, and mean peak consumption for all households during the pre-trial period (half a year) was 0.828 kWh.
	Therefore these treatment effects are of the order of 10\% of peak consumption.

	Below we present two estimates of single causal trees as an example of the instability of single tree estimates and small sample size. Causal forest Individual Treatment Effect (\textsc{ite}) estimates are then plotted with confidence intervals and summarized in density plots. We also apply the methods described by \cite{chernozhukov2017generic} to test the hypothesis $\beta_2=0$ in equation (\ref{BLP_eq}), and confirm that there is heterogeneity of treatment effects and that Causal forest Individual Treatment Effect (\textsc{ite}) estimates are relevant predictors of the true \textsc{ite}s. We test the association between Causal forest Individual Treatment Effect (\textsc{ite}) estimates and a set of pre-trial variables using the approach of \cite{chernozhukov2017generic}. Finally, variable importance measures are presented in order to consider which variables are the strongest determinants of the structure of the trees in the forest.


\subsection*{Causal Trees}
	
\begin{figure}
	\includegraphics[width=0.9\textwidth]{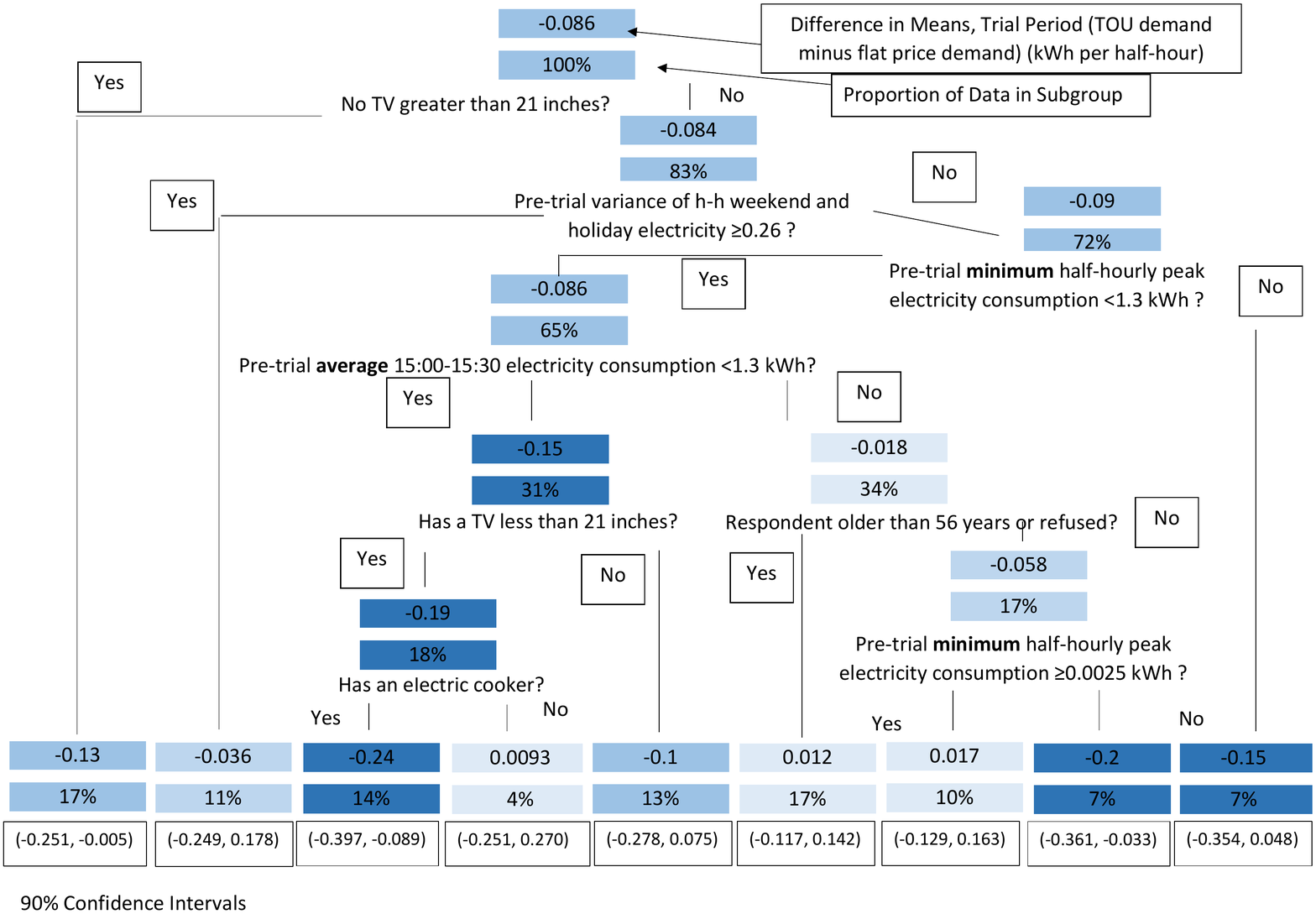}
	\begin{minipage}[t]{\textwidth}
		\caption{Single Tree Example 1}
		\label{Tree_example1}
	\end{minipage}
\end{figure}

\begin{figure}
	\includegraphics[width=0.9\textwidth]{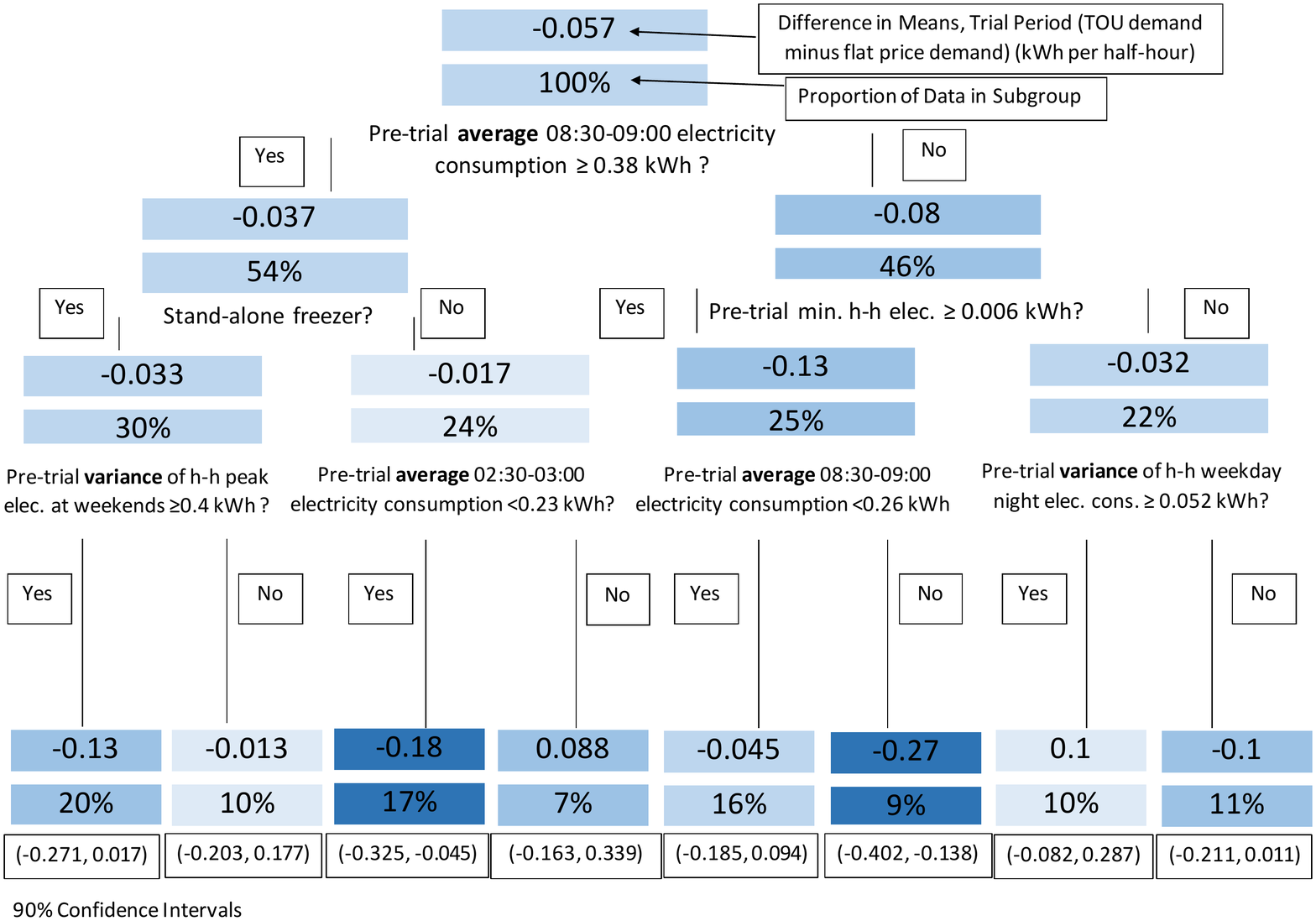}
	\begin{minipage}[t]{\textwidth}
		\caption{Single Tree Example 2 - Different seed}
		\label{Tree_example2}
	\end{minipage}	
\end{figure}

Figures \ref{Tree_example1} and \ref{Tree_example2} show estimated causal trees.\footnote{The trees were obtained using the \pkg{R} package \pkg{causalTree}.} 
The set of potential splitting variables is given in Table \ref{fig: potential_split_vars_DH_surv_usage}.\footnote{The trees are ``honest'', i.e. separate data is used for obtaining splitting points and for obtaining terminal node estimates. Half of the data is used for creating the splits in the tree, and half is used for honest estimation. The minimum number of treatment and control observations required for a leaf split is set to ten. } 
The only difference in estimation of the two trees is the seed for random number generation.\footnote{The seed determines the subsampling of the data into splitting and estimation data, and determines subsamples used for cross-validation.} The diagrams contain 90\% confidence intervals.

It can be immediately observed from these trees that the partition of the data generated by the causal tree algorithm is sensitive to the input data. This can be viewed as partly a sample size issue. Sample size, in combination with sample splitting for honest estimation, also has implications for statistical significance.  There were 500 observations used for splitting, and 501 observations for estimation of treatment effects. The causal tree output contains few subgroups with significantly non-zero treatment effects at the 5\% level. In contrast, \textsc{cate} estimates obtained from a low-variance method, such as a linear model interacting treatment with different levels of education and including control variables, can result in multiple groups with significant effects. 

The above instability   can be addressed by the use of a causal forest. The instability of the output (i.e. sensitivity to the random separation of the data into splitting and estimation subsamples) is less of a problem when aggregation of predictions occurs over a large number of honest causal trees. 


		\subsection*{Causal Forest}
	
We fitted a causal forest to the dataset containing a set of control households and households allocated to tariff \textsc{c} and the \textsc{ihd} stimulus (1001 households).\footnote{The causal forest algorithm was implemented using the \pkg{R} package \pkg{grf}.} 
Each individual honest tree is fitted using a bootstrap sample consisting of half of the data, with half of this sample used for splitting and half used for estimation.\footnote{\cite{bertrandcontemporaneous} also use these sizes of bootstrap samples and training and estimation subsamples. \cite{wager2018estimation} divide bootstrap samples in half for honest estimation.} The number of individual trees fitted is 15000.\footnote{This is somewhat arbitrary, and between the values of 10000 and 25000 used by \cite{bertrandcontemporaneous} and \cite{davis2017using}.} For each tree in the forest, a random subsample of one third of the set of covariates are used as potential splitting variables.  \footnote{The choice of one third of the total number of covariates is commonly used for random forests.} The minimum number of leaf observations is set to the default of five
.


First, we applied the \textsc{blp} test of \cite{chernozhukov2017generic} to test for heterogeneity, as outlined in the methods section. The results are presented in Table \ref{BLP_table1}. The test strongly rejects the null hypothesis that $\beta_2=0$, suggesting that there is heterogeneity in demand response and the causal forest \textsc{ite} estimate is a relevant predictor, i.e. the \textsc{ite} estimates have a non-zero coefficient when interacted with the treatment indicator variable. A comparison with the results obtained from other machine learning methods in Table \ref{BLP_table_others} suggests that the casual forest \textsc{ite} estimates are more relevant linear predictors of the true \textsc{ite}s than the estimates produced by the other methods.

Table \ref{GATEs_table} provides a further description of demand response heterogeneity. The average peak demand reduction per half-hour is -0.150kWh for the 25\% of households that reduce demand the most, while average demand reduction is not significantly nonzero for the 25\% least responsive households.

Tables \ref{CLAN_table_usage} and \ref{CLAN_table_survey} suggest that our estimates provide a reasonable characterisation of heterogeneity.\footnote{For some variables of interest, particularly binary variables with few non-zero values, confidence interval could not be obtained because for some sample splits there was not sufficient variation within quartiles for a confidence interval to be calculated.} In Table \ref{CLAN_table_usage} we test for differences in averages of pre-trial electricity consumption variables between the 25\% of households with the highest and lowest demand response. Unsurprisingly, the most responsive households consume significantly more, and have significantly more variable consumption than the least responsive households. Table \ref{CLAN_table_usage} contains tests for differences in averages of binary survey variables. For example, we can observe that for the first quartile of treatment effects, i.e. the quartile of most responsive households, 40.5\% of households have a respondent with third level education.


For the vast majority of covariates we observe associations across quantiles of individual effects that we would expect a priori. The most responsive households (i.e. Quartile 1) generally use more electricity, are more educated, younger, higher social class, and have more appliances. This particular result is in agreement with the observation made by \cite{dicosmolyons2014}, using the same data, that more educated households are generally more responsive.\footnote{Our focus on peak demand response is also justified by the observation by \cite{di2017nudging} that households ``reduced consumption rather than shifting consumption from peak''.}

Tables \ref{CLAN_table_age}, \ref{CLAN_table_class}, and \ref{CLAN_table_employment} demonstrate that demographic groups that are more likely to contain vulnerable customers \citep{CSE2012}, namely lower class and retired households, together with households for which the respondent was over 65 years old, contain a greater proportion of less responsive households. While this may be largely due to the fact that these groups have less reducible peak usage, this difference in demand response for vulnerable and non-vulnerable groups could be relevant to regulation of potential consumer targeting. 
The patterns of heterogeneity observed in both Tables \ref{CLAN_table_usage} and \ref{CLAN_table_survey} are mostly maintained when the forest is fitted using only electricity consumption data.\footnote{The results for causal forests fitted using only survey variables or only usage variables are not included in this article, but are available from the authors on request.}

\begin{figure}
		\includegraphics{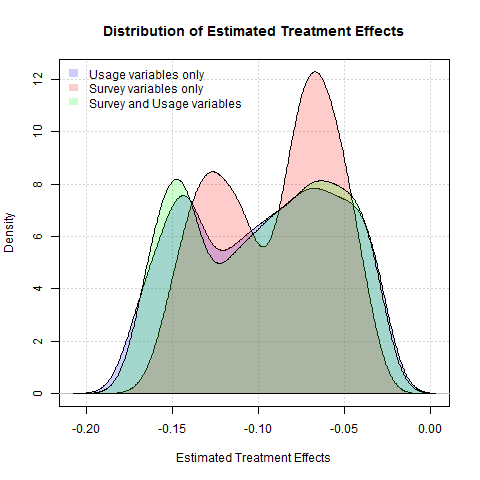}
		\caption{Standard splitting}
		\label{density_ITEs}
	\caption{Density plots of causal forest household estimates fitted using different sets of variables}
\end{figure}

	
	  To demonstrate this, Figure \ref{density_ITEs} presents a density plot comparing the distributions of the \textsc{ite} estimates obtained by fitting causal forests with different sets of potential conditioning variables. One forest was fitted using both survey and usage variables, one forest was fitted using only usage variables, and one forest was fitted using only survey variables. This suggests that electricity consumption data contains information related to survey data information that can characterise heterogeneous groups of demand response. This issue may be relevant to firms or policymakers who wish to understand which information to collect in order to predict demand response.

	The results suggest that the usage variables exert a greater influence on the causal forest estimates. Furthermore, the density plot suggests potential bimodality in the distribution of individual effects. However, although it is most plausible that past usage variables are more informative than survey variables, we also note the possibility that these results are driven by the bias of variable selection towards continuous variables, which have more potential splitting points. This issue can be addressed by discretizing each continuous usage variable, for example, into indicator variables defined by quantiles. 

\FloatBarrier
			
\subsection*{Variable Importance}
	
In this section we present the results for variable importance utilising the methods outlined in Section \ref{section_methods}.
The variable importance measure is a depth-weighted average of the number of splits on the variable of interest.\footnote{This is the default measure in the \pkg{R} package \pkg{grf}.} For the second method we also carry out a permutation-based test, as outlined in Section \ref{section_methods}.

Columns (1) and (2) of Table \ref{varimptable} give the variable names and values for the variable importance measure. The variables are ordered by importance, with larger values indicating greater importance. The importances are scaled such that the most important variable has variable importance equal to 100.

The results indicate that the trees most often split on electricity usage, and specifically variables that indicate the level and variance of weekday electricity consumption. The most important survey variables are number of laptop PCs, number of freezers, and employment status. These variables are likely to be correlated with income and level of electricity usage.



As noted in Section \ref{section_methods}, given the bias of variable importance measures in favour of variables with more splitting points \citep{strobl2008statistical}, we implement an alternative \emph{permutation} test of variable importance which is able to address this issue \citep{altmann2010permutation}. Column (3) shows the p-values for the permutation tests on the \pkg{grf} variable importances. 
The p-values confirm the pattern of results observed in column (2). 

	

\FloatBarrier

\section{Conclusion}\label{conclusion}
	
In this article we have examined  heterogeneity of demand response following the introduction of time-of-use electricity pricing. 
Variable importance measures, adjusted for differences in information content across past usage and demographic variables,  suggest that the causal forest algorithm  favours the use of certain functions of past electricity consumption  rather than survey information to describe heterogeneity. 
Tables \ref{CLAN_table_survey} to \ref{CLAN_table_employment} reveal notable patterns of heterogeneity across \emph{unimportant} survey variables. For example, the causal forest results suggest that younger, more educated households that consume more electricity exhibit greater demand response to new pricing schemes.  In this respect, although survey variables can be less informative than detailed electricity consumption information in terms of selection in the causal forest algorithm, they can also be correlated with \emph{important} past consumption information.

\pagebreak
	
	\bibliographystyle{apa}
	\bibliography{First_Chapter_refs}

\pagebreak
	
\newpage	

\newpage

\newpage

\newpage

\newpage	
	
\begin{figure}	
	\includegraphics[width=1\textwidth]{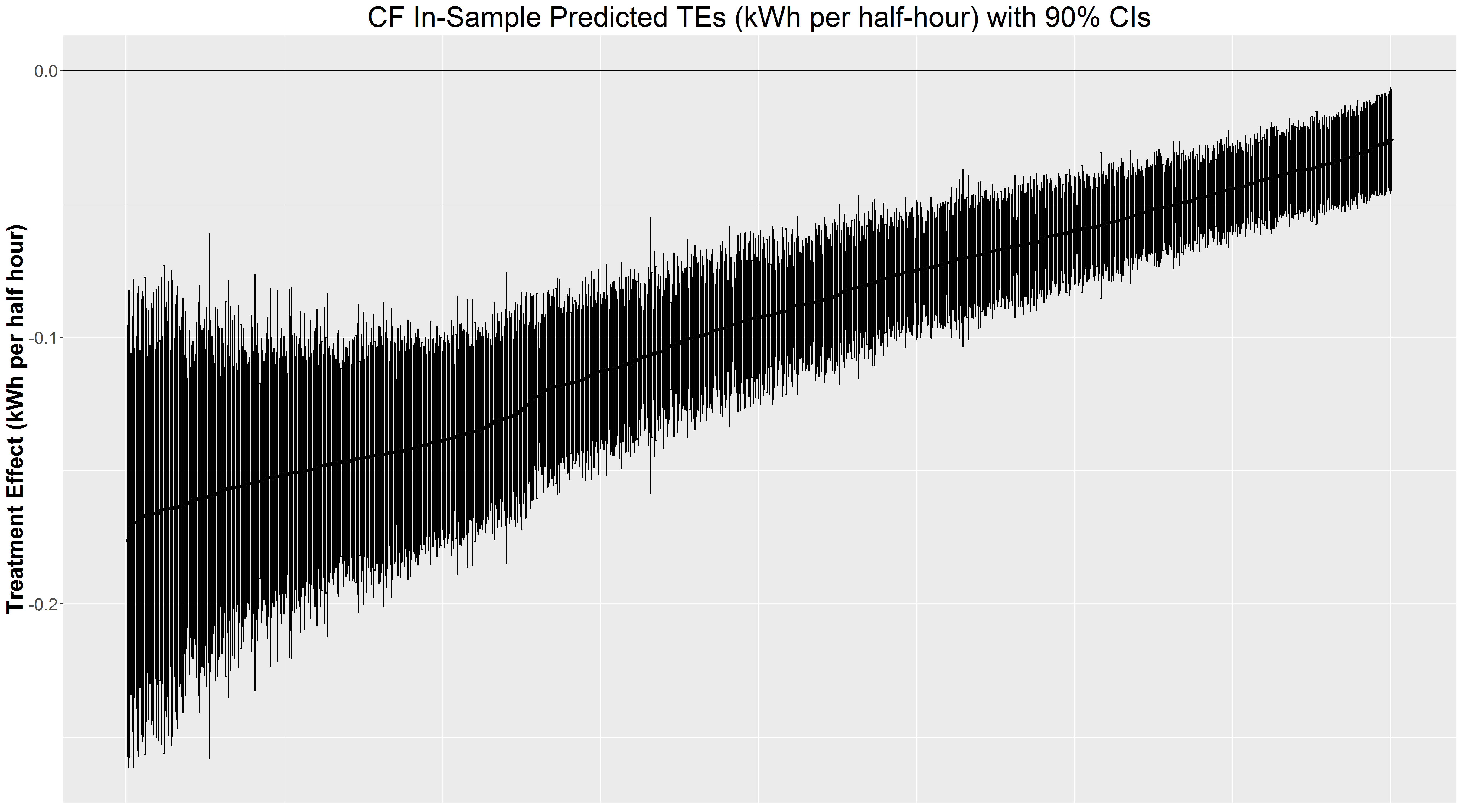}
	\caption{90\% Confidence Intervals for ITEs ordered by size of ITE estimate}
	\label{Wager_CIs_ordered}
\end{figure}

\newpage

\begin{table}[!htbp] \centering
	\caption{\textsc{tou} Tariff details}
	\label{tariff_table}
	\begin{tabular}{llll }
		\hline
		\textbf{\textsc{TOU} Tariffs} & \textbf{Night}& \textbf{Day}& \textbf{Peak}\\
		(cents per kWh)& 23.00-08.00 & 08.00-17.00 every day & 17.00-19.00 Mon-Fri\\
		& & 19.00-23.00 every day & Excluding holidays \\	
		& & 17.00-19.00 weekends &  \\
		& & and holidays & \\	
		\hline
		Tariff A & 12.00 & 14.00 & 20.00 \\
		Tariff B & 11.00 & 13.50 & 26.00\\
		Tariff C & 10.00 & 13.00 & 32.00\\	
		Tariff D & 9.00 & 12.50 & 38.00\\	
	\end{tabular}
\end{table}

\newpage

\begin{table}[!htbp] \centering
	{\small
		\caption{Potential splitting variables for Causal Trees and Causal Forest}
		\label{fig: potential_split_vars_DH_surv_usage}
		\begin{tabular}{ll}
			\\[-1.8ex]\hline
			\hline \\[-1.8ex]
			\textbf{Name of variable}	& 	\\		
			\hline
			
			\textbf{Survey variables (categorical)}	& 	\\
			Age of respondent  	& Sex of respondent	\\
			Class of chief income earner			& Regular internet use	\\	
			Employment status of chief income earner			& Other reg. internet users	\\								
			Number of bedrooms		 		& Education of chief earner	\\							
			Type of home				&	Electric central heating \\							
			Alone or other occupants			& Electric plugin heating	\\								
			Own or rent the home			& Central water heating	\\								
			Number of electric cookers - number			& Immersion water heating	\\								
			Internet access				& Instant water heating	\\							
			Approximate age of home				& Number of washing machines	\\															
			Lack money for heating			& Number of tumble dryers	\\	
			Number of dishwashers		& Number of instant electric showers	\\															
			No. showers elec. pumped from hot tank			& Type of cooker	\\															
			Number of plug-in convector heaters			& Number of freezers	\\																													
			Number of water pumps or electric wells			& Number of immersion water heaters	\\																													
			Number of small TVs			& Number of big TVs	\\												
			Number of desktop PCs			& Number of laptop PCs	\\														
			Number of games consoles			& Has an energy rating	\\														
			Proportion of energy saving lightbulbs			& Prop. double glazed windows	\\	
			Lagging jacket 			& Attic insulation\\																																																																																				
			External walls insulated			& 	 \\
			
			& \\															
			\textbf{Electricity usage variables (continuous)}	&  \\															
			Mean usage  &  Min. usage\\															
			Variance of usage	 & Max. usage \\							
			Mean peak usage & Mean nonpeak usage \\										
			Variance of peak usage & Variance of nonpeak usage \\	
			Mean night usage & Mean daytime usage \\										
			Variance of night usage & Variance of daytime usage \\		
			Mean usage - weekdays & Mean peak usage - weekdays \\															
			Variance of usage - weekdays	& Var. peak usage - weekdays \\	
			Mean night usage - weekdays & Mean daytime usage - weekdays \\										
			Variance of night usage - weekdays & Var. daytime usage - weekdays \\																		
			Mean daily maximum usage  & Mean usage - weekends \\										
			Mean daily minimum usage  & Variance of usage - weekends \\										
			Mean of half-hour coefficients of variation & Mean usage - each month (July-Dec) \\													
			Avg. night usage/ avg. daily usage & Var. of usage - each month (July-Dec) \\										
			Avg. lunchtime usage/ Avg. daily usage  & Mean usage - each half-hour \\
			Mean night usage - weekends & Mean daytime usage - weekends \\										
			Variance of night usage - weekends & Var. daytime usage - weekends \\												
			\hline
		\end{tabular}
		
	}
\end{table}

\newpage

\begin{table}[!htbp] \centering
	\caption{BLP of Peak Demand}
	\label{BLP_table1}
	\begin{tabular}{L{3.5cm}C{3.5cm}C{3.5cm} }
		\\[-1.8ex]\hline
		\hline \\[-1.8ex]
		& \multicolumn{2}{l}{\textit{}} \\
		
		Variable & ATE($\beta_1$) & HET($\beta_2$)  \\
		
		\hline \\[-1.8ex]
		Half-hourly peak consumption (kWh) & -0.095 &  1.620 \\
		& (-0.128,-0.062)   & (0.636,2.600)  \\		  						
		\hline
		\multicolumn{3}{l}{\footnotesize{Medians over 1000 splits. 90\% confidence interval in parenthesis}} \\
		
	\end{tabular}
	
\end{table}

\newpage

\begin{table}[!htbp] \centering
	\caption{BLP of Peak Demand}
	\label{BLP_table_others}
	\begin{tabular}{L{3.5cm}|C{2.5cm}C{2.5cm}|C{2.5cm}C{2.5cm} }
		\\[-1.8ex]\hline
		\hline \\[-1.8ex]
		& \multicolumn{2}{c}{\textit{Elastic Net}} &\multicolumn{2}{c}{\textit{Boosted Tree}}\\
		
		Variable & ATE($\beta_1$) & HET($\beta_2$)   & ATE($\beta_1$) & HET($\beta_2$)\\
		
		\hline \\[-1.8ex]
		Half-hourly peak consumption (kWh) & -0.010 & 0.486 & -0.098 & 0.189 \\
		& (-0.135,-0.067)   & (0.214,0.762) & (-0.131,-0.064) &  (-0.081,0.463) \\	
		\hline \\[-1.8ex]
				\hline \\[-1.8ex]
		& \multicolumn{2}{c}{\textit{Neural Network}}& \multicolumn{2}{c}{\textit{Random Forest}}\\	  		
		Variable & ATE($\beta_1$) & HET($\beta_2$)   & ATE($\beta_1$) & HET($\beta_2$) \\				
		\hline
		Half-hourly peak consumption (kWh) & -0.093 & 0.035 & -0.097 & 0.364 \\
		& (-0.131,-0.056) & (-0.124,0.195) & (-0.129,-0.065) & (0.026,0.707) \\	
		
		\multicolumn{5}{l}{\footnotesize{Medians over 1000 splits. 90\% confidence interval in parenthesis}} \\
		\multicolumn{5}{l}{\footnotesize{The ML methods were implemented in \pkg {R} using the package \pkg{caret} and method names \pkg{glmnet}, \pkg{gbm}, \pkg{pcaNNet}, and \pkg{rf}}} \\
	\end{tabular}
	
\end{table}

\newpage

	\begin{table}[!htbp] \centering
		\caption{Group Average Treatment Effects (GATEs) for most and least peak demand responsive households}
		\label{GATEs_table}
		\begin{tabular}{L{3.5cm}C{3.5cm}C{3.5cm}C{3.5cm} }
			\\[-1.8ex]\hline
			\hline \\[-1.8ex]
			& \multicolumn{3}{l}{\textit{}} \\
			
			Variable & 25\% most responsive & 25\% least responsive & Difference  \\
			
			\hline \\[-1.8ex]
			Half-hourly peak consumption (kWh) & -0.150 &  -0.046 & 0.105 \\
						  & (-0.202, -0.098)   & (-0.097, 0.006) &  (0.028, 0.181) \\		  						
			\hline
			 \multicolumn{4}{l}{\footnotesize{Medians over 1000 splits. 90\% confidence interval in parenthesis}} \\
			
		\end{tabular}

	\end{table}

\newpage	
		
\begin{table}[!htbp] \centering
	\caption{Classification Analysis (CLAN): Pre-trial electricity consumption variable averages for most and least peak demand responsive households}
	\label{CLAN_table_usage}
	\begin{tabular}{L{7cm}C{3cm}C{3cm}C{2.5cm} }
		\\[-1.8ex]\hline
		\hline \\[-1.8ex]
		& \multicolumn{3}{l}{\textit{}} \\
		
		Variable & 25\% most responsive & 25\% least responsive & Difference  \\
		
		\hline \\[-1.8ex]
				 &  &  & \\
		Avg. pre-trial half-hourly usage (kWh) & 0.804 & 0.229 & 0.574 \\
		 & (0.771,0.835)  & (0.216,0.242) & (0.540,0.609) \\
		 				 &  &  & \\
		Avg. pre-trial peak half-hourly usage (kWh) & 1.412 & 0.344 & 1.068 \\
				 & (1.358,1.467) & (0.321,0.367) & (1.009,1.127) \\
					 &  &  & \\
				Var. of pre-trial half-hourly usage (kWh)  &  0.779 & 0.109 & 0.669   \\
				 & (0.722,0.833) & (0.097,0.121) & (0.613,0.725)\\
			 &  &  & \\
			 Var. pre-trial peak half-hourly usage (kWh) & 1.307 & 0.168 & 1.139 \\						
				 & (1.215,1.402) & (0.146,0.190) & (1.042, 1.236) \\
			 &  &  & \\
			 Max half-hour elec. con. (kWh) & 7.688 &  3.828 & 3.862 \\ 			
				 & (7.414,7.960) & (3.601,4.055) & (3.508,4.217) \\
			 &  &  & \\
			 Min half-hour elec. cons. (kWh) & 0.037  & 0.013 & 0.025 \\ 			
				 & (0.028,0.047) & (0.010,0.016) & (0.014,0.035) \\
			 &  &  & \\
			 Mean daily max (kWh)  & 3.607 & 1.295 & 2.309 \\ 			
				 & (3.489,3.723) & (1.212,1.377) & (2.168,2.451) \\
			 &  &  & \\
			 Mean daily min (kWh)  & 0.149 & 0.042 & 0.107 \\ 			
				 & (0.133,0.165) & (0.037,0.048) & (0.090,0.123) \\
		\hline

	\end{tabular}
\end{table}

\newpage	

\begin{table}[!htbp] \centering
	\caption{Classification Analysis (CLAN): Survey variable averages for most and least peak demand responsive households}
	\label{CLAN_table_survey}
	\begin{tabular}{L{7cm}C{3cm}C{3cm}C{2.5cm} }
		\\[-1.8ex]\hline
		\hline \\[-1.8ex]
		& \multicolumn{3}{l}{\textit{}} \\
		
		Variable & 25\% most responsive & 25\% least responsive & Difference  \\
		
		\hline \\[-1.8ex]
		&  &  & \\
		Male & 0.516 & 0.488 & 0.020\\
		& (0.427,0.604)  & (0.399,0.577) & (-0.144,0.105) \\
		&  &  & \\
		Internet access & 0.873 & 0.424 & 0.457 \\
		& (0.814,0.932) & (0.336,0.512) & (0.359,0.559) \\
		&  &  & \\
		Electric central heating &  0.032 & 0.048 & -0.016   \\
		& & (0.010,0.086) & (-0.033,0.064) \\
		&  &  & \\
		Water immersion & 0.635 & 0.424 & 0.211  \\						
		& (0.550,0.720) & (0.336,0.512) & (0.089,0.333) \\
		&  &  & \\
		Water centrally heated & 0.159 &  0.112 & 0.047 \\ 			
		& (0.094,0.223) & (0.056,0.168) &  (-0.041,0.128) \\
		&  &  & \\
		Went without heat from lack of money & 0.048  & 0.040 & 0.000 \\ 			
		& (0.010,0.085) & (0.005,0.075) & (-0.053,0.048) \\
		&  &  & \\
		Lagging jacket on hot water  & 0.857 & 0.776 &  0.081 \\ 			
		& (0.795,0.919) & (0.702,0.850)  &  (-0.017,0.177)\\
		&  &  & \\
		Third level education  & 0.405  &  0.288  &  0.125 \\ 			
		& (0.318,0.492) & (0.208,0.368) & (0.005,0.241) \\
		&  &  & \\
		Employee & 0.563  & 0.328   & 0.235  \\ 			
		& (0.476,0.651) & (0.245,0.411) & (0.114,0.355) \\
		&  &  & \\
		Apartment & 0  & 0.048   & -0.048  \\ 			
		&  & (0.010,0.086) &  (-0.086,-0.01) \\
		&  &  & \\
		Instantaneous water heater &  0.008 & 0.024   & -0.016  \\ 			
		&  &  &  \\
		&  &  & \\
		Plug-in electric heater & 0.032  &  0.024  & 0.008  \\ 			
		&  &  & (-0.047,0.033) \\
		&  &  & \\

		\hline
		\multicolumn{4}{L{\textwidth}}{\footnotesize{Note: For some binary variables with few non-zero values, confidence interval could not be obtained because for some sample splits there was not sufficient variation within quartiles for a confidence interval to be calculated.}} \\
		
	\end{tabular}

\end{table}

\newpage	

\begin{table}[!htbp] \centering
	\caption{Classification Analysis (CLAN): Age variable averages for most and least peak demand responsive households}
	\label{CLAN_table_age}
	\begin{tabular}{L{7cm}C{3cm}C{3cm}C{2.5cm} }
		\\[-1.8ex]\hline
		\hline \\[-1.8ex]
		& \multicolumn{3}{l}{\textit{}} \\
		
		Variable & 25\% most responsive & 25\% least responsive & Difference  \\
		
		\hline \\[-1.8ex]
		&  &  & \\
		18-25 & 0 & 0 & 0\\
		&   & &  \\
		&  &  & \\
		26-35 & 0.095 & 0.080 & -0.015 \\
		& (0.043,0.147) & (0.032,0.128) & (-0.083,0.058) \\
		&  &  & \\
		36-45 & 0.230  & 0.152 &  -0.078  \\
		& (0.156,0.305) & (0.088,0.216) & (-0.178,0.017) \\
		&  &  & \\
		46-55 & 0.349 & 0.168 &  -0.181 \\						
		& (0.265,0.434) & (0.102,0.234) & (-0.288,-0.074,) \\
		&  &  & \\
		56-65 & 0.206 & 0.192  & -0.014 \\ 			
		& (0.135,0.278) & (0.122,0.262) &  (-0.115,0.084,) \\
		&  &  & \\
		65+ & 0.103  & 0.392 & 0.289 \\ 			
		& (0.049,0.157) & (0.305,0.479) & (0.190,0.392,) \\
		&  &  & \\
		Refused  & 0.016 & 0.008 & -0.008  \\ 			
		&  &   & \\
		&  &  & \\		
		\hline
		\multicolumn{4}{L{\textwidth}}{\footnotesize{Note: For some binary variables with few non-zero values, confidence interval could not be obtained because for some sample splits there was not sufficient variation within quartiles for a confidence interval to be calculated.}} \\
		
	\end{tabular}
\end{table}

\newpage	

\begin{table}[!htbp] \centering
	\caption{Classification Analysis (CLAN): Class variable averages for most and least peak demand responsive households}
	\label{CLAN_table_class}
	\begin{tabular}{L{7cm}C{3cm}C{3cm}C{2.5cm} }
		\\[-1.8ex]\hline
		\hline \\[-1.8ex]
		& \multicolumn{3}{l}{\textit{}} \\
		
		Variable & 25\% most responsive & 25\% least responsive & Difference  \\
		
		\hline \\[-1.8ex]
		&  &  & \\
		Upper middle and middle & 0.190 & 0.056 & 0.135 \\
		& (0.121,0.260)  &  & (0.057,0.217) \\
		&  &  & \\
		Lower middle & 0.294 & 0.216 & 0.070 \\
		& (0.213,0.374) & (0.143,0.289) & (-0.037,0.181) \\
		&  &  & \\
		Skilled working & 0.190  & 0.128 &  0.055  \\
		& (0.121,0.260) & (0.069,0.187) & (-0.032,0.148) \\
		&  &  & \\
		Working and non-working & 0.302 & 0.560 & -0.258  \\						
		& (0.220,0.383) & (0.472,0.648) & (-0.377,-0.139) \\
		&  &  & \\
		Farmers & 0.016 & 0.024  & 0.008 \\ 			
		&  &  &   \\
		&  &  & \\
		Refused & 0.008  & 0.008 & 0.000 \\ 			
		&  &  &  \\
		&  &  & \\
		\hline

	\end{tabular}
\end{table}

\newpage	

\begin{table}[!htbp] \centering
	\caption{Classification Analysis (CLAN): Employment variable averages for most and least peak demand responsive households}
	\label{CLAN_table_employment}
	\begin{tabular}{L{7cm}C{3cm}C{3cm}C{2.5cm} }
		\\[-1.8ex]\hline
		\hline \\[-1.8ex]
		& \multicolumn{3}{l}{\textit{}} \\
		
		Variable & 25\% most responsive & 25\% least responsive & Difference  \\
		
		\hline \\[-1.8ex]
		&  &  & \\
		Employee & 0.0563 & 0.328 & 0.235\\
		& (0.0476,0.0651)  & (0.245,0.411) & (0.113,0.355,) \\
		&  &  & \\
		Self-emp (with employees) & 0.095 & 0.008 & 0.087 \\
		& (0.043,0.147) &  & (0.032,0.141) \\
		&  &  & \\
		Self-emp (with no employees) & 0.071  & 0.032 &  0.039  \\
		&(0.026,0.117) &  & (-0.015,0.097) \\
		&  &  & \\
		Unemployed (seeking work) & 0.048 & 0.072 & -0.024  \\						
		& (0.010,0.085) & (0.026,0.118) & (-0.084,0.035) \\
		&  &  & \\
		Unemployed (not seeking work) & 0.040 & 0.040  & -0.008 \\ 			
		&  &  &  (-0.055,0.043,) \\
		&  &  & \\
		Retired &  0.159 & 0.496 & -0.337 \\ 			
		& (0.094,0.223) & (0.497,0.585) & (-0.445,-0.227) \\
		&  &  & \\
		Carer  & 0.024 & 0.008 &  0.008 \\ 			
		&  &   &  (-0.022,0.038)\\
		&  &  & \\		
		\hline
		\multicolumn{4}{L{\textwidth}}{\footnotesize{Note: For some binary variables with few non-zero values, confidence interval could not be obtained because for some sample splits there was not sufficient variation within quartiles for a confidence interval to be calculated.}} \\
		
	\end{tabular}
\end{table}

\newpage	
	
\begin{table}[ht]
	\centering
		\resizebox{!}{0.45\textheight}{
	\begin{tabular}{lll|lll}
		\hline
		Variable name & Variable importance & p-value & Variable name & Variable importance & p-value\\
		\hline
\textit{electric plugin heating} & 0 & 0.09 & mean 14:00-14:30 usage & 12.23 & 0.79 \\
\textit{water instantly heated} & 0 & 0 & mean usage - weekdays & 12.9 & 0.1 \\
\textit{unheated, lack of money} & 0 & 0.38 & var. night usage - weekends & 12.95 & 0.99 \\
\textit{number of washing machines }& 0.02 & 0.47  & mean daytime usage - weekends & 13 & 0.14 \\
\textit{electric central heating} & 0.22 & 0.9  & mean night / mean day usage & 13.07 & 1 \\
\textit{prop. double glazed windows} & 0.25 & 1  & mean 21:30-22:00 usage & 13.07 & 0.73 \\
\textit{number of electric cookers} & 0.46 & 1  & mean 22:30-23:00 usage & 13.28 & 0.9 \\
\textit{number of immersion heaters} & 0.54 & 1  & var. night usage - weekdays & 13.35 & 1 \\
\textit{number of dishwashers} & 0.58 & 1  & mean 13:00-13:30 usage & 13.72 & 0.84 \\
\textit{type of cooker} & 0.84 & 1  & mean 06:30-07:00 usage & 14.14 & 0.98 \\
\textit{number of tumble dryers} & 0.85 & 1  & mean daytime usage - weekdays & 14.87 & 0.16 \\
\textit{water centrally heated} & 1.06 & 1  & mean 14:30-15:00 usage & 15.23 & 0.71 \\
\textit{regular internet user} & 1.07 & 1  & mean usage - weekends & 15.3 & 0.07 \\
\textit{sex of respondent} & 1.23 & 1  & var. daytime usage - weekdays & 15.36 & 0.4 \\
\textit{own or rent home} & 1.24 & 1  & mean 19:00-19:30 usage & 15.36 & 0.61 \\
\textit{no. of elec. convector heaters} & 1.33 & 1  & mean 21:00-21:30 usage & 15.49 & 0.49 \\
\textit{water pumped from elec. well} & 1.58 & 0.99  & mean 07:30-08:00 usage & 15.96 & 1 \\
\textit{attic insulated} & 1.79 & 1  & mean h-h coef. of variation & 16.6 & 1 \\
\textit{number of instant elec. showers} & 2.03 & 1  & var. nonpeak usage - weekdays & 16.64 & 0.26 \\
\textit{external walls insulated} & 2.06 & 1  & mean 23:00-23:30 usage & 16.76 & 0.85 \\
\textit{other internet users} & 2.07 & 0.49  & \textit{number of freezers} & 17 & 0.06 \\
\textit{water immersion} & 2.22 & 0.98  & variance nonpeak usage & 17.12 & 0.12 \\
\textit{number of small TVs} & 2.35 & 1  & mean 00:00-00:30 usage & 17.39 & 0.84 \\
\textit{number of hot tank elec. showers} & 2.36 & 0.97   & \textit{number of laptop PCs} & 17.48 & 0.1 \\
\textit{age of home} & 2.92 & 1  & variance daytime usage & 18.06 & 0.16 \\
\textit{number of games consoles} & 2.94 & 0.89  & mean 10:00-10:30 usage & 19.76 & 0.51 \\
\textit{education} & 3.44 & 1 & mean 20:00-20:30 usage & 19.93 & 0.12 \\
\textit{lagging jacket} & 3.45 & 0.44  & variance of usage & 20.1 & 0.05 \\
\textit{has an energy rating} & 3.46 & 0.44  & mean daily min. usage & 20.72 & 0.72 \\
\textit{age of respondent }& 3.91 & 1  & mean 23:30-00:00 usage & 21.12 & 0.68 \\
\textit{prop. elec. saving lightbulbs} & 4.74 & 1  & mean 18:00-18:30 usage & 21.29 & 0.25 \\
\textit{number of bedrooms} & 4.87 & 0.98  & var. usage - weekends & 21.41 & 0.13 \\
\textit{lives alone} & 5.17 & 0.69  & mean 18:30-19:00 usage & 21.82 & 0.19 \\
mean 06:00-06:30 usage & 5.61 & 1  & mean 19:30-20:00 usage & 22.01 & 0.17 \\
mean 02:30-03:00 usage & 5.66 & 1  & var. usage - weekdays & 22.11 & 0.05 \\
\textit{type of home} & 5.78 & 0.97  & var. daytime usage - weekends & 22.85 & 0.1 \\
mean 04:00-04:30 usage & 6.03 & 1  & mean 16:00-16:30 usage & 22.86 & 0.46 \\
mean 12:00-12:30 usage & 6.1 & 1  & mean 09:00-09:30 usage & 23.05 & 0.58 \\
\textit{internet access} & 6.35 & 0.12  & mean November peak usage & 24.88 & 0.15 \\
mean 03:00-03:30 usage & 6.44 & 1  & min. half-hourly usage & 25.8 & 0.72 \\
mean 03:30-04:00 usage & 6.68 & 1  & mean 16:30-17:00 usage & 26.02 & 0.51 \\
mean 05:00-05:30 usage & 6.73 & 1  & var. November peak usage & 26.19 & 0.39 \\
mean night usage & 6.76 & 0.99  & max. half-hourly usage & 26.96 & 0.67 \\
mean 04:30-05:00 usage & 7.12 & 1  & mean lunchtime / mean day usage & 27.47 & 0.92 \\
\textit{number of big TVs} & 7.42 & 0.68  & mean 15:30-16:00 usage & 29.14 & 0.18 \\
mean 11:00-11:30 usage & 7.5 & 1  & mean daily max. usage & 29.39 & 0.06 \\
mean night usage - weekends & 7.53 & 0.99  & mean 15:00-15:30 usage & 30.3 & 0.13 \\
mean 05:30-06:00 usage & 7.87 & 1  & mean 08:00-08:30 usage & 30.85 & 0.48 \\
mean 11:30-12:00 usage & 8.22 & 1  & mean 20:30-21:00 usage & 31.45 & 0.04 \\
mean 00:30-01:00 usage & 8.48 & 1  & var. December peak usage & 38.92 & 0.2 \\
\textit{social class} & 8.95 & 0.64  & mean 09:30-10:00 usage & 39.05 & 0.13 \\
mean 01:30-02:00 usage & 9.02 & 0.99  & mean 08:30-09:00 usage & 39.44 & 0.21 \\
\textit{number of desktop PCs} & 9.11 & 0.12  & mean peak usage - weekdays & 42.61 & 0 \\
mean 12:30-13:00 usage & 9.36 & 1  & mean peak usage & 44.2 & 0 \\
mean night usage - weekdays & 9.38 & 0.96  & variance peak usage & 44.65 & 0.01 \\
mean 13:30-14:00 usage & 9.83 & 0.99  & var. peak usage - weekdays & 47.83 & 0.03 \\
mean 01:00-01:30 usage & 10.31 & 0.98  & mean July peak usage & 49.55 & 0.08 \\
mean nonpeak usage - weekdays & 10.31 & 0.3  & mean September peak usage & 49.96 & 0.03 \\
variance night usage & 10.47 & 1  & mean 17:00-17:30 usage & 57.62 & 0.02 \\
mean 02:00-02:30 usage & 10.78 & 0.94  & mean December peak usage & 62.11 & 0 \\
mean 10:30-11:00 usage & 10.86 & 0.99  & var. September peak usage & 66.87 & 0.03 \\
\textit{employment} & 11.22 & 0.4  & var. July peak usage & 68.34 & 0.1 \\
mean nonpeak usage & 11.66 & 0.17  & mean August peak usage & 68.73 & 0 \\
mean of usage & 11.86 & 0.1  & mean 17:30-18:00 usage & 69.41 & 0 \\
mean 07:00-07:30 usage & 11.9 & 1  & var. August peak usage & 74.89 & 0.01 \\
mean 22:00-22:30 usage & 11.96 & 0.85  & mean October peak usage & 76.14 & 0 \\
mean daytime usage & 12.2 & 0.19  & var. October peak usage & 100 & 0 \\

				\hline
	\end{tabular}
}
	\raggedright Survey variables are in italics.
\caption{Variable Importance results}
\label{varimptable}
\end{table}

\FloatBarrier

\newpage

\begin{appendices}
\section{}\label{simstudy_perm_test}

\textbf{Simulation Study - Variable Importance Permutation Test}

We present a simulation study investigating the extent to which p-values for a permutation-based variable importance test are influenced by the bias of the variable importance measure towards continuous variables and categorical variables with more categories. This study is designed in a similar way to that used by \cite{strobl2008statistical} for investigating the bias of random forest variable importance measures.

First, we generate the following covariates and treatment indicator: $X_1 \sim N(0,1) , \ X_2 \sim Cat(2), \ X_3 \sim Cat(4),\ X_4 \sim Cat(10),\ X_5 \sim Cat(20),\ treatment \sim Cat(2)$, where $Cat(k)$ denotes a categorical distribution with $k$ categories of equal probability. Then we consider simulations of the outcome under the following three model designs:

For design 1, none of the covariates affect the outcome, and the outcome is normally distributed: $Y \sim N(0,1)$
For design 2 and 3, the dependent variable is defined in a similar way to a simulation study carried out by \cite{athey2016recursive}:
\begin{equation}
Y = \eta(X)+\frac{1}{2}(2 \times treatment -1) \times  \kappa(X) +\epsilon
\end{equation}
where $\epsilon \sim N(0,1)$ . For design 2 the functions are $\eta(X)=0$ and $\ \kappa(X)=X_2$, and for design 3 the functions are $\eta(X)=\frac{1}{2}X_1+X_2$ and $\ \kappa(X)=X_2$.

\begin{figure}[ht] 
	\centering
	\begin{subfigure}{0.32\textwidth}
		\includegraphics[trim={0 0 0 2cm},clip,width=\linewidth]{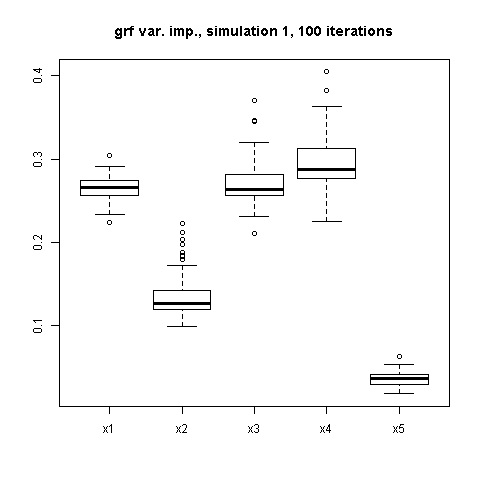}
		\caption{Design 1 var. imp.}
		\label{grf_imp_design1}
	\end{subfigure}
	\begin{subfigure}{0.32\textwidth}
		\includegraphics[trim={0 0 0 2cm},clip,width=\linewidth]{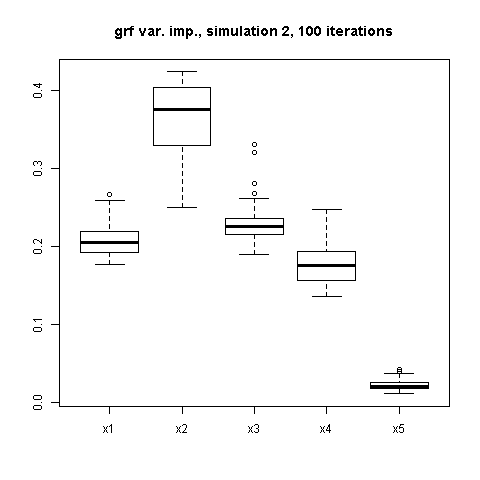}
		\caption{Design 2 var. imp.}
		\label{grf_imp_design2}
	\end{subfigure}
	\begin{subfigure}{0.32\textwidth}
		\includegraphics[trim={0 0 0 2cm},clip,width=\linewidth]{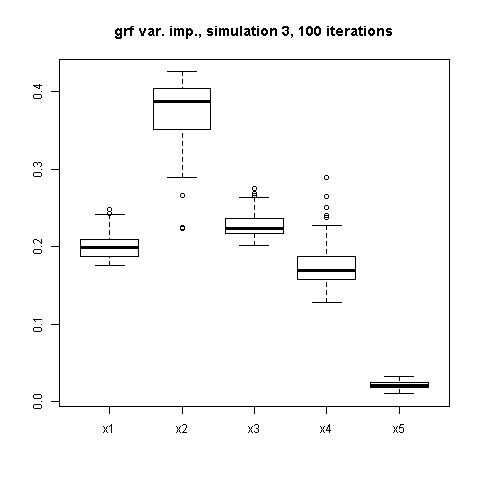}
		\caption{Design 3 var. imp.}
		\label{grf_imp_design3}
	\end{subfigure}
	\caption{Boxplots of simulation study variable importances, 100 permutations, 100 iterations} \label{boxplots_simulation_varimps}
\end{figure}

\begin{figure}[ht] 
	\centering
	\begin{subfigure}{0.32\textwidth}
		\includegraphics[trim={0 0 0 2cm},clip,width=\linewidth]{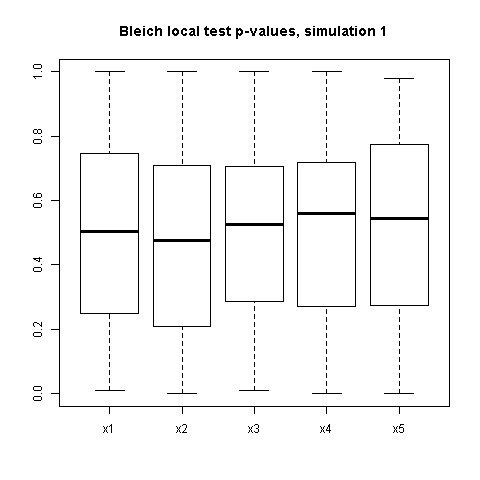}
		\caption{Design 1 p-values}
		\label{pval_design1}
	\end{subfigure}
	\begin{subfigure}{0.32\textwidth}
		\includegraphics[trim={0 0 0 2cm},clip,width=\linewidth]{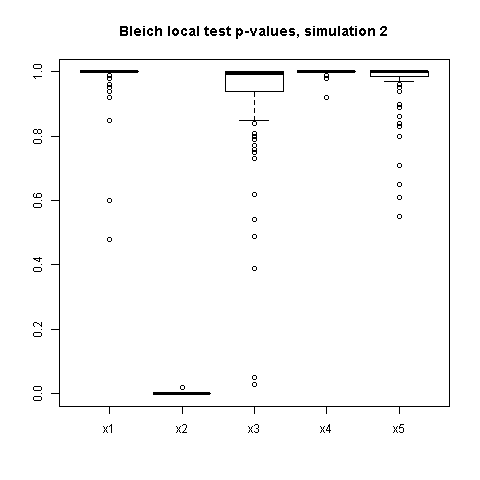}
		\caption{Design 2 p-values}
		\label{pval_design2}
	\end{subfigure}
	\begin{subfigure}{0.32\textwidth}
		\includegraphics[trim={0 0 0 2cm},clip,width=\linewidth]{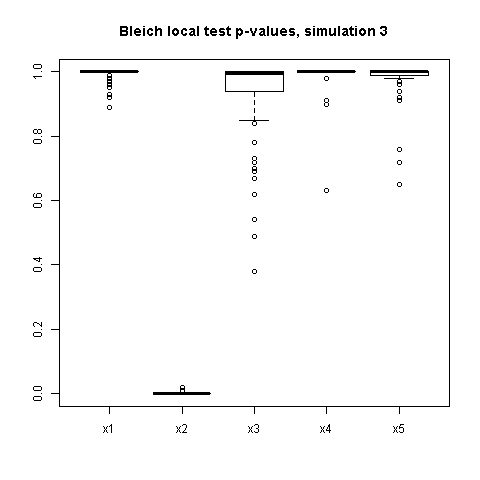}
		\caption{Design 3 p-values}
		\label{pval_design3}
	\end{subfigure}	
	\caption{Boxplots of simulation study p-values, 100 permutations, 100 iterations} \label{boxplots_simulation_pvals}
\end{figure}

We simulate these designs 100 times, with 500 observations per simulation, and for each simulation we permute the dependent variable 100 times and obtain p-values, and then present boxplots of the p-values for each variable.\footnote{The parameters for the causal forest are: Number of trees = 5000, bootstrap sample fraction = 0.5, number of potential splitting variables random selected at each split = number of variables divided by 3 and rounded down, minimum node size = 5.} The boxplots of variable importances obtained using the unpermuted dependent variable are shown in Figure \ref{boxplots_simulation_varimps}. The boxplots for the p-values are shown in Figure \ref{boxplots_simulation_pvals}. 
The boxes give the lower quartile, median, and upper quartiles across repeated simulations. The whiskers give the most extreme data points that are no more than $1.5$ times the interquartile range from the box. The circles denote outliers.

Note that the results in Figures \ref{boxplots_simulation_varimps} and \ref{boxplots_simulation_pvals} should be interpreted differently. The variable importances in Figure \ref{boxplots_simulation_varimps} are not used in a test of significance, but rather in a comparison of importance across variables. In contrast, Figure \ref{boxplots_simulation_pvals} is clearer and correctly indicates that the binary variable is significant in designs 2 and 3. This is an argument in favour of the permutation test.

Although for  design 1 none of the variables affects the outcome, in Figure \ref{grf_imp_design1} $X_1$ has greater variable importance than $X_2$, because of the aforementioned bias towards continuous variables.

For categorical variables $X_3$, $X_4$, and $X_5$, all with more categories than $X_2$, there are two factors influencing the bias of the variable importance measure. As the number of categories increases, there are more potential splits on the variable of interest, because there is a binary variable for each category. This explains why $X_3$ has greater variable importance than $X_2$ in Figure \ref{grf_imp_design1}. On the other hand, considering the case of a variable with a large number of categories, $X_5$, there will be relatively few observations allocated to any one category, and therefore a split on one of the $X_5$ categories is unlikely to lead to a large improvement in the splitting criterion. Therefore the variable importance measures for $X_5$ are small.

The p-values in Figure \ref{boxplots_simulation_pvals} appear to be unaffected by these biases. In Figure \ref{pval_design1}, none of the variables tend to have significant p-values, reflecting the fact that none of the variables has any influence on the outcome.

In Figures \ref{pval_design2} and \ref{pval_design3}, $X_2$ is correctly identified as the important variable. Although Figures \ref{grf_imp_design2} and \ref{grf_imp_design3} also indicate that $X_2$ is the most important variable, there are also misleading differences in the importances of the other variables. However, in Figures \ref{pval_design2} and \ref{pval_design3}, the variables $X_1$, $X_3$, $X_4$, and $X_5$ tend to have similar, insignificant p-values.

\FloatBarrier

\FloatBarrier

\end{appendices}

\end{document}